\pdfoutput=1

\documentclass[aps,prb,twocolumn,showpacs,groupedaddress,floatfix]{revtex4}
\usepackage{graphicx}

\begin{document}

\title{A phenomenological model for the pressure sensitivity of the Curie temperature in
hole-doped manganites}

\author{G. Garbarino}
\thanks{Present address: ID27-ESRF, BP 166 cedex 09, 38042 Grenoble, France.}
\author{C. Acha}
\thanks {Also fellow of CONICET of Argentina}
\email{acha@df.uba.ar} \affiliation{Laboratorio de Bajas
Temperaturas, Departamento de F\'{\i}sica, FCEyN, Universidad
Nacional de Buenos Aires, Pabell\'on I, Ciudad Universitaria,
C1428EHA Buenos Aires, Argentina}

\date{\today}


\begin{abstract}

We performed high pressure experiments on
La$_{0.8}$Ca$_{0.2-x}$Sr$_x$MnO$_3$ (LCSMO)($0\leq x \leq0.2$)
ceramic samples in order to analyze the validity of the well known
relation between the A mean ionic radius ($\langle r_A \rangle$) and
the Curie temperature $T_c$  of hole-doped manganites at a fixed
doping level and for doping values below the 0.3
(Mn$^{+4}$/Mn$^{+3}$) ratio. By considering our results and
collecting others from the literature, we were able to propose a
phenomenological law that considers the systematic dependence of
$T_c$ with structural and electronic parameters. This law predicts
fairly well the pressure sensitivity of $T_c$, its dependence with
the A-cation radius disorder and its evolution in the high pressure
range. Considering a Double Exchange model, modified by polaronic
effects, the phenomenological law obtained for $T_c$ can be
associated with the product of two terms: the polaronic modified
bandwidth and an effective hole doping.

\end{abstract}

\maketitle

\section{INTRODUCTION}

Many efforts have been devoted to determine the relevant electronic
and structural parameters that fix the Curie temperature (T$_c$) of
manganites~\cite{Hwang95a,Radaelli97,Dagotto01}. The Double Exchange
model~\cite{Zener51,Anderson55} (DE) was initially applied in order
to correlate electrical transport properties and magnetic ordering
in these compounds. But early experiments such as the temperature
dependence of the Hall coefficient~\cite{Jaime97}, the differences
in the activation energy between thermopower and
conductivity~\cite{Jaime96}, and the isotope
effect~\cite{Babushkina98} demonstrated the polaronic nature of the
carriers. This evidence showed the necessity for introducing
polaronic corrections to the electronic bandwidth that determines
$T_c$.

On the other hand, it was experimentally established that, for
hole-doped manganites and particularly for the family
A$_{0.7}^{'}$A$^{''}_{0.3}$MnO$_{3}$ (where A$^{'}$ is a trivalent
rare earth ion and A$^{''}$ a divalent alkali earth ion), the
resulting A mean ionic radius ($\langle r_A \rangle$) has a clear
influence on $T_c$.~\cite{Hwang95b} Experiments also showed that
$\langle r_A \rangle$ can be varied both by chemical replacement or
by an external pressure~\cite{Hwang95b}, where in the former case,
both the Mn-O bond distance and the Mn-O-Mn bond angle vary while in
the latter case, most of the variation comes from the Mn-O bond
distance.~\cite{Radaelli97} Although the $T_c(\langle r_A \rangle)$
dependence can be well reproduced by varying the pressure just by
considering a linear dependence $\delta \langle r_A \rangle = \gamma
\delta P$ ($\gamma = 3.75 10^{-4} {\AA}$/kbar) for $P \leq$ 20 kbar.
In this low pressure range, $T_c$ varies linearly with $P$, while
for higher pressures, $T_c$ reaches a maximum value and decreases
for a further increase of $P$.~\cite{Neumeier95,Manolo01} This
behavior seems to be related to pressure-dependent competing
interactions, like the ferromagnetic (F) and the antiferromagnetic
(AF) coupling between the core spins, as suggested by Sacchetti et
al. ~\cite{Sacchetti06}, although other factors that govern the
polaronic modified DE model can play a major role. On the other
hand, Rivadulla et al.~\cite{Rivadulla06} made a good quantitative
description of the $\langle r_A \rangle$ dependence of $T_c$ at
constant doping based on a mean field model where the reduction of
the volume fraction of the itinerant electrons produced by the phase
separation is responsible for the observed behavior.

Empirically, the temperature dependence of the pressure ($P$)
sensitivity of T$_c$ (dlnT$_c$/dP) was established, which seems to
represent an universal behavior for many moderated hole-doped
manganites~\cite{Moritomo97}. This curve could be described
qualitatively within the small polaron modified DE model but it was
far from quantitative~\cite{Laukhin97} and even considering polarons
in the more suitable intermediate electron-phonon coupling regime
did not produce a better understanding.~\cite{Lorenz01}

In this paper we present a phenomenological model based on the
$\langle r_A \rangle$ dependence of T$_c$ for intermediate to large
bandwidth AMnO$_3$ hole-doped perovskites which usefully describes
the quantitative dependence of T$_c(P)$. A DE interaction, modified
by polaronic effects and also an effective doping of the MnO planes,
both controlled by $\langle r_A \rangle$ are suggested as the two
microscopic ingredients that govern the proposed relation.

\section{EXPERIMENTAL}

Previous experiments~\cite{Hwang95b} showed that $T_c$ follows a
parabolic dependence with $\langle r_A \rangle$  for
La$_{1-y}$T$_y$MnO$_3$ (T=Sr; Ca; Pr) for a fixed doping level $y
\sim$ 0.3. Here, in order to test the validity of this dependence
for other doping levels ($y=0.2$), we performed resistivity
measurements as a function of temperature and pressure on
La$_{0.8}$Ca$_{0.2-x}$Sr$_x$MnO$_3$ (LCSMO)($0\leq x \leq0.2$)
ceramic samples. These samples were synthezised following a similar
process to the one published elsewhere.\cite{Ulyanov01} The
temperature dependence of resistivity was measured using a
conventional 4 terminal DC technique in a CuBe pinston-cylinder
hydrostatic cell described previously.\cite{Garbarino06} Pressures
up to 10 kbar were applied using a 50 \% mixture of kerosene and
transformer oil as the pressure transmitting medium. Pressure was
measured at room temperature by using a calibrated InSb sensor and
it remains constant over all the temperature range (within a 10\% of
variation for a temperature span of 77 K to 350 K) in spite of
thermal contractions. Temperature was measured using a calibrated
carbon-glass thermometer in good thermal contact with the cell's
body.

\section{RESULTS AND DISCUSSION}

\begin{figure} [b]
\includegraphics[width=3in]{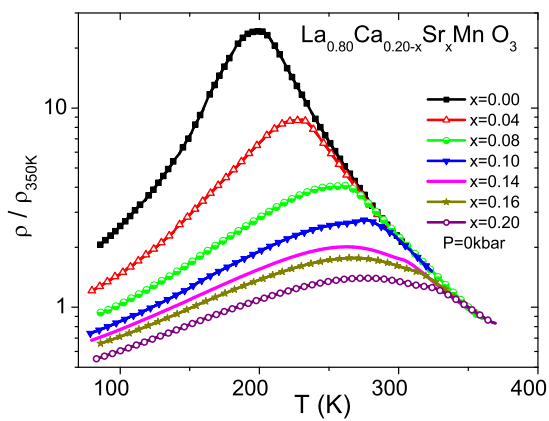}
\vspace{-0mm} \caption{(Color online) Temperature dependence of the
normalized resistivity of La$_{0.8}$Ca$_{0.2-x}$Sr$_x$MnO$_3$
($0\leq x \leq0.2$)} \vspace{5mm} \label{fig:RvsTyX}
\end{figure}

The normalized resistivity as a function of temperature of the LCSMO
series with $0\leq x \leq0.2$ and $y$=0.2 can be observed in
Fig.~\ref{fig:RvsTyX}. All the curves show a change in the
conduction regime that can be associated with a metal to insulator
transition ($T_{MI}$), which increases with increasing Sr content.

\begin{figure}
\includegraphics[width=3.0in]{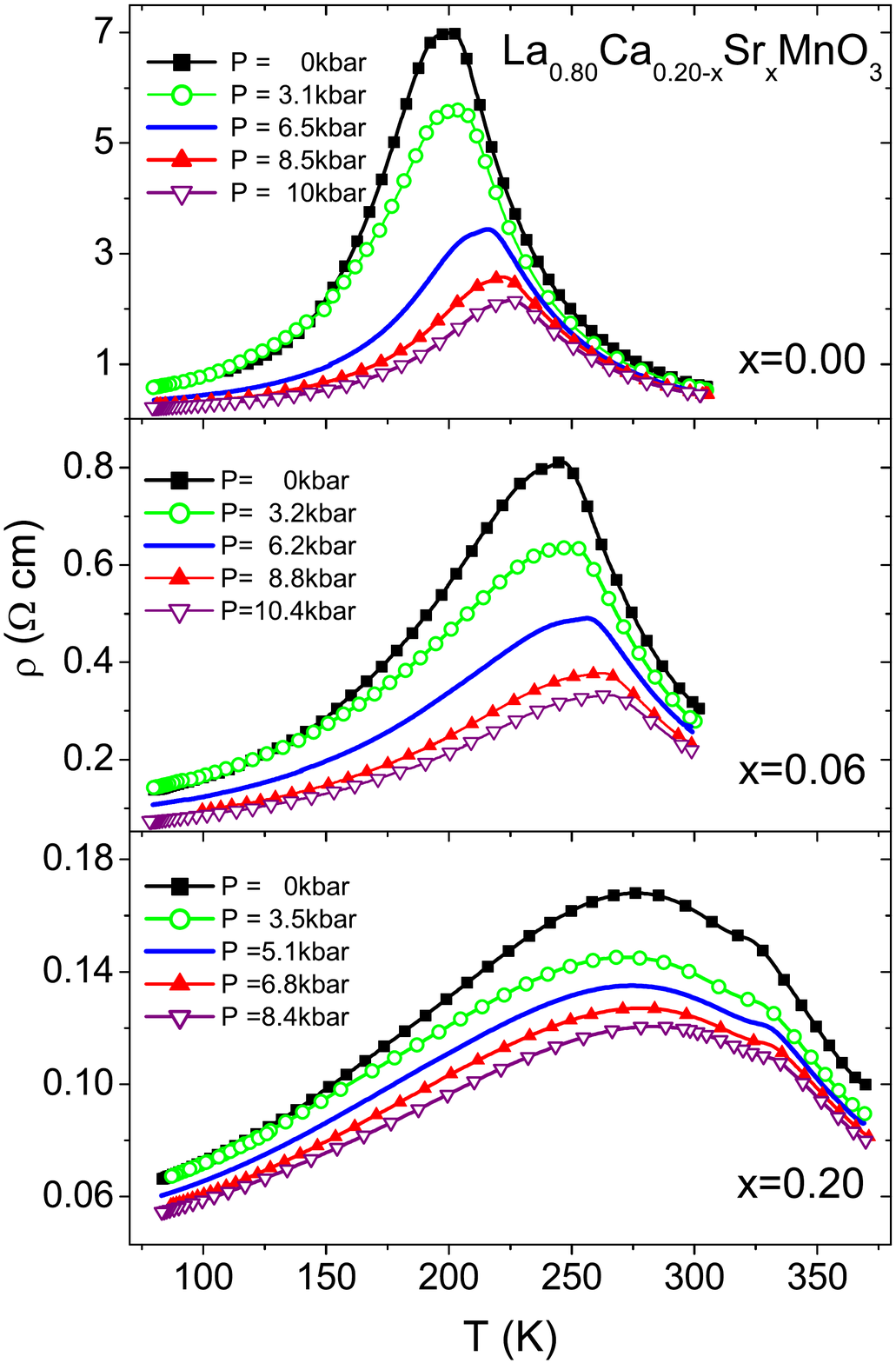}
\vspace{-0mm} \caption{(Color online) Pressure sensitivity of the
resistivity as a function of temperature of
La$_{0.8}$Ca$_{0.2-x}$Sr$_x$MnO$_3$ ($x$=0; 0.06; 0.2).}
\vspace{5mm} \label{fig:RvsTyP}
\end{figure}

The pressure sensitivity of the resistivity is shown in
Fig.~\ref{fig:RvsTyP} for samples LCSMO with $x$=0; 0.06 and 0.20.
Pressure increases both $T_{MI}$ and the conductivity of these
materials.

\begin{figure}
\includegraphics[width=3.3in]{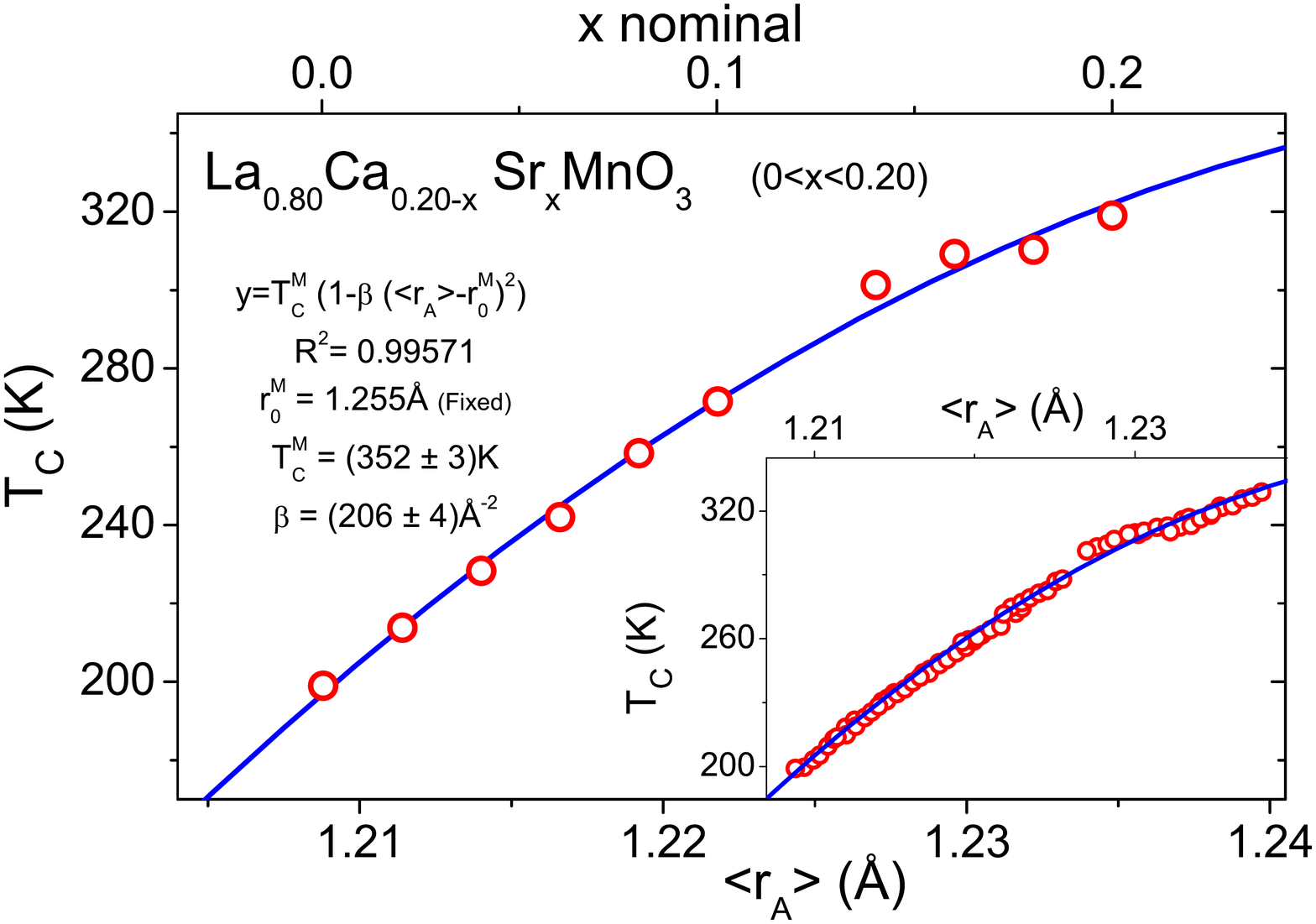}
\vspace{-0mm} \caption{(Color online) $T_c$ vs the Sr concentration
($x$) or the average ionic radius of the cation in the A site
($\langle r_A \rangle$) at constant doping $y=0.2$. The line is a
fit using Eq. 1; the fitting parameters are displayed. The inset
also shows $T_c$ vs $\langle r_A \rangle$, but where the variation
of $\langle r_A \rangle$ is due to both chemical replacements and
external pressure (assuming that $\delta T_c^M / dP = \delta \beta /
dP =0$ and $\delta \langle r_A \rangle= \gamma \delta P$, with
$\gamma=3.75~10^{-4}$ \AA/kbar).} \vspace{5mm} \label{fig:TcvsRa}
\end{figure}

$T_{MI}$ does not necessarily coincide with the Curie temperature,
so we determine $T_c$ from the resistivity curves as the temperature
at which a sudden increase in the logarithmic temperature derivative
of the resistivity is observed. It has been shown previously that
this coincides with the $T_c$ determined by magnetization
measurements.~\cite{Cui04} By following this criteria and by
calculating the variation on $\langle r_A \rangle$ generated by
chemical replacements (from Shannon´s tables of ionic
radii~\cite{Shannon76}) or by external pressure (assuming that
$\gamma$ is independent of the doping level and that $\langle r_A
\rangle$ is the only pressure-dependent parameter), we obtain the
dependence of $T_c$ as a function of $\langle r_A \rangle$ for the
LCSMO ($y=0.2$) samples, shown in Fig.~\ref{fig:TcvsRa}.

\clearpage
The data is very well represented by a quadratic law for
the whole pressure and doping intervals considered, which indicates
that the assumptions we made were quite reasonable. A small
departure from the ideal dependence can be observed for $\langle r_A
\rangle \simeq$ 1.227 \AA, which coincides with a structural
transition reported for this series.\cite{Ulyanov02}

\begin{figure} [b]
\includegraphics[width=3.3in]{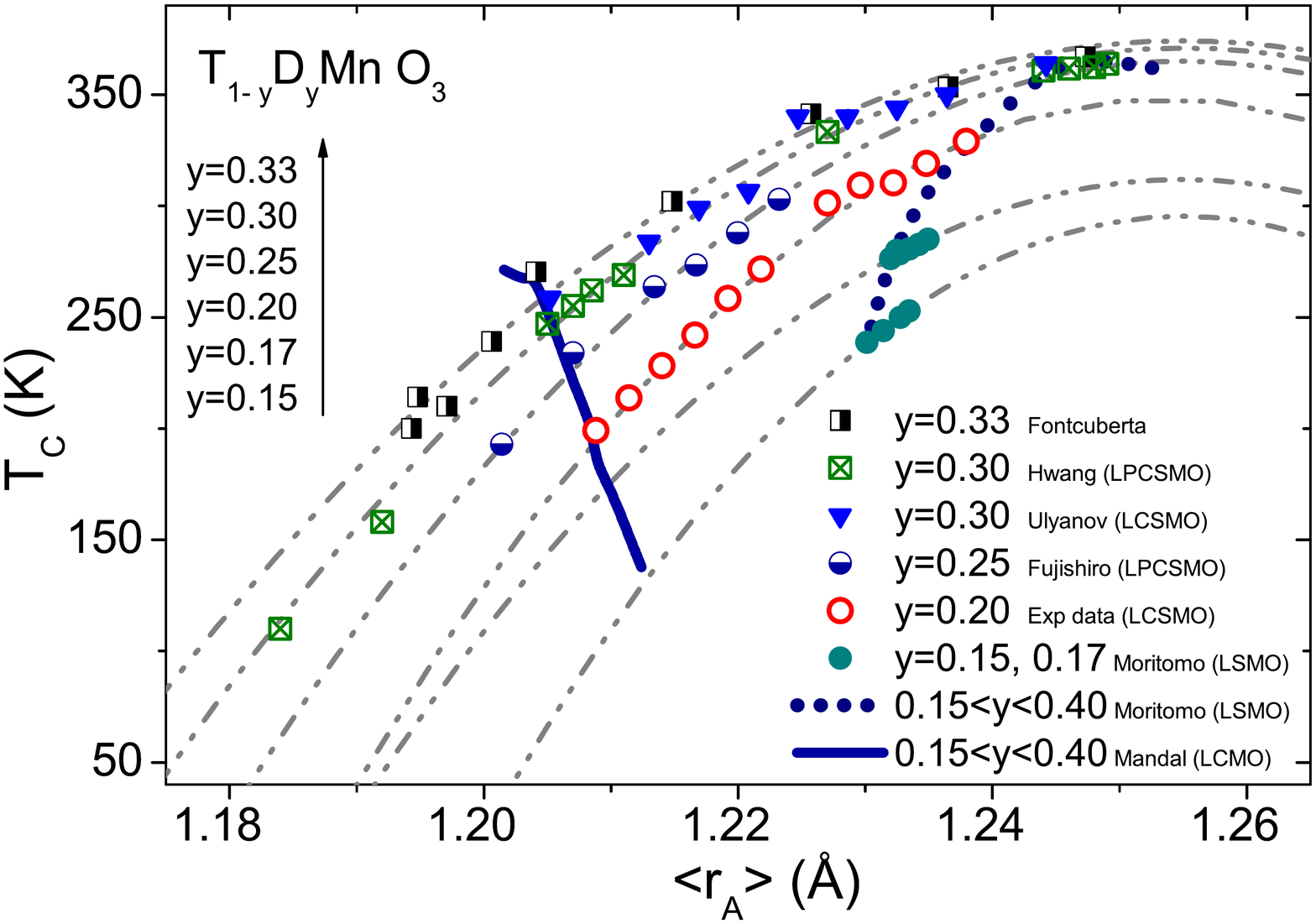}
\vspace{-0mm} \caption{(Color online) The phase diagram of
T$_{1-y}$D$_y$MnO$_3$ (where T is a trivalent lanthanide as La, Sm,
Nd, and D a divalent alkaline earth as Ca, Sr) for 0.15 $\leq$ y
$\leq$ 0.33 as a function of $\langle r_A \rangle$. Data was
extracted from references\cite{Fontcuberta98,Hwang95b,Ulyanov02,
Fujishiro00,Moritomo95,Mandal03} The dashed lines are parabolic fits
corresponding to Eq.~\ref{eq:Tcvsra} at constant doping $y$. The
evolution of $T_c$ and $\langle r_A \rangle$ is also shown for the
Sr and Ca-doped LMO samples (i.e. samples with a varying $y$).}
\vspace{5mm} \label{fig:TcvsRagral}
\end{figure}

From our data and the data already published we can extend the study
of the $T_c(\langle r_A \rangle$) dependence for other manganites
and for doping levels $y$ in the 0.15 $\leq$ $y \leq$ 0.33 range.
The obtained $T_c(y,\langle r_A \rangle$) curves, shown in
Fig.~\ref{fig:TcvsRagral}, follow the same general behavior: a
parabolic law for each doping concentration and a doping-dependent
maximum [$T_c^M(y)$] located at $\langle r_A^M \rangle \simeq$ 1.255
\AA. This dependence can be represented by an expression of the form

\begin{equation}
\label{eq:Tcvsra} T_c(y,\langle r_A \rangle) = T_c^M(y)\left[1-
\beta (y)\left(\langle r_A \rangle(x,y) - \langle r_A^M
\rangle\right)^2\right],
\end{equation}

\begin{figure}
\includegraphics[width=3.3in]{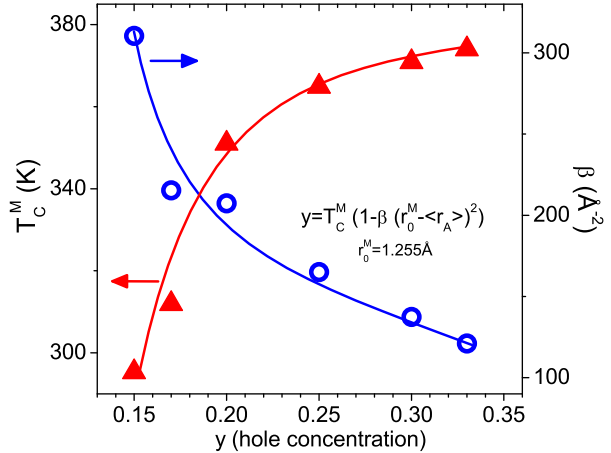}
\vspace{-0mm} \caption{(Color online) The fitted parameters $T_c^M$
and $\beta$ as a function of the hole concentration $y$. Lines are
guides to the eye.} \vspace{0mm} \label{fig:param}
\end{figure}

By fitting the data presented in Fig.~\ref{fig:TcvsRagral} we can
determine the doping sensitivity of parameters $T_c^M$ and $\beta$,
as can be observed in Fig.~\ref{fig:param}. As was shown in a
previous study~\cite{Attfield96}, the local structural disorder,
generated by the occupation of the $A$ site by cations with
different sizes, produces a reduction of the ideal $T_c$ that would
be measured in case that this disorder does not exist. The disorder
can be quantified by the variance $\sigma^2$ of the A-cation radius
distribution. In order to perform the fits, minimizing in this way
the contribution of disorder in the obtained parameters, we choose
from the $T_c(y,\langle r_A \rangle,\sigma$) data the points with
small $\sigma^2$ ($<10^{-3}$ \AA$^2$). On the other hand, if we
assume that Eq.~\ref{eq:Tcvsra} gives the $T_c$ of a manganite with
negligible A-cation radius disorder [$T_c^{\star}(x,\langle r_A
\rangle)=T_c(x,\langle r_A \rangle,\sigma=0)$], we can estimate the
$T_c$ of a manganite with a structural disorder $\sigma$ in $r_A$ as
the mean $T_c$ resulting from a uniform distribution of cells with
A-cation radii within the interval $r_A \pm \sigma$. The result
gives an expression of the form

\begin{equation}
\label{eq:Tcprom} \langle T_c[y,\langle r_A \rangle,\sigma^2]\rangle
= T_c^{\star}(y,\langle r_A \rangle) - (T_c^M(y)
\beta(y)/3)\sigma^2,
\end{equation}

which gives a simple explanation of the linear dependence of $T_c$
on $\sigma^2$ already published for the perovskite family
A$^{'}_{0.7}$A$^{''}_{0.3}$MnO$_3$.~\cite{Attfield96} The
$\sigma^2$'s pre-factor can be calculated from the fitted parameters
shown in Fig.~\ref{fig:param}. For $y=0.3$ we obtain a value of
(17.000 $\pm$ 1000) K\AA$^{-2}$, quite similar to the experimental
data published. Also, we can qualitatively estimate the influence of
the disorder on the pressure sensitivity of $T_c$ by taking the
pressure derivative of Eq.~\ref{eq:Tcprom}. If we compare the data
of previous papers ~\cite{Manolo01,Cui03} it is clear that
$d\sigma^2/dP$ is an increasing function of $\sigma$, the second
term of the right part of the derived equation would indicate a
reduction of the expected pressure sensitivity of manganite with
increasing $\sigma^2$, as was experimentally obtained
previously.\cite{Otero07} Besides, considering the similarity of Eq.
(1) with the one developed by Bean and Rodbell~\cite{Bean62} to
describe the coupling of magnetic order to structural distortions, a
first order magnetic phase transition at $T_c$ can be predicted for
large values of $\beta$, as was demonstrated experimentally by
Otero-Leal et al..~\cite{Otero07}

The expression of the pressure sensitivity of $T_c$, shown in
Eq.~\ref{eq:dlnTcdP2}, can be easily obtained from
Eq.~\ref{eq:Tcvsra} as,

\begin{equation}
\label{eq:dlnTcdP2} \frac{dlnTc}{dP} =  \frac{dlnT_c^M}{dP}  + 2
\gamma \sqrt{\beta}
\sqrt{\frac{T_c^M}{T_c}\left(\frac{T_c^M}{T_c}-1\right)},
\end{equation}

\begin{figure}
\includegraphics[width=3.1in]{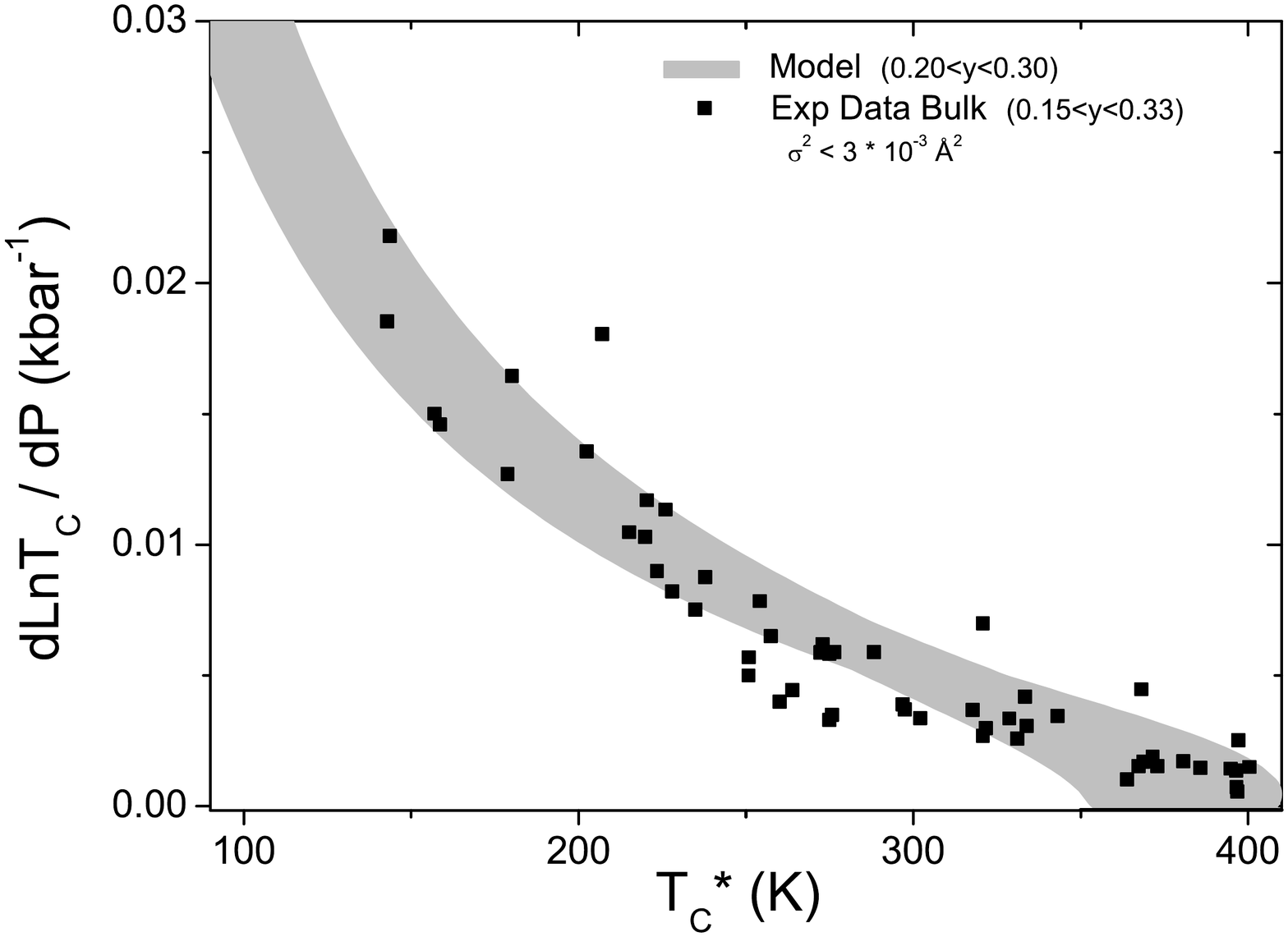}
\vspace{-0mm} \caption{Pressure sensitivity as a function of
T$_c^{\star}$ for compounds of the La$_{1-y}$T$_y$MnO$_3$ family
(T=Sr; Ca; Y, Dy)  in the 0.15 $\leq y \leq$ 0.33 range. To
guarantee the range of validity of Eq.~\ref{eq:Tcprom} in order to
estimate T$_c^{\star}$, data points with $\sigma < 3~10^{-3}
{\AA}^2$ were extracted from references
\cite{Hwang95b,Fontcuberta98,Moritomo95,Laukhin97,Manolo01,Neumeier95,
Deteresa96,Postorino03,Ulyanov02,Lorenz01} and from our
measurements. The shaded area represents the predictions of the
model (Eq.\ref{eq:dlnTcdP2}) taking into account the different
values of the fitted parameters ($\beta$ and $T_c^{M}$).}
\vspace{5mm} \label{fig:dlnTcvsTestr}
\end{figure}

By using Eq.~\ref{eq:dlnTcdP2} and the fitted parameters $T_c^M(y)$
and $\beta(y)$, the pressure sensitivity of $T_c$ at low pressures
can be predicted for many compounds. A good accordance between
experimental points and the predicted behavior, represented by a
shaded area as we considered the doping dependence of the fitted
parameters, can be observed in Fig.~\ref{fig:dlnTcvsTestr}. Here, we
included data points where $T_c$ was determined by different
criteria and techniques (ac susceptibility or resistivity) which
accounts for the dispersion of data. We only applied the restriction
that the selected data points should be derived from compounds with
a small structural disorder in $r_A$ ($\sigma <$ 3 10$^{-2} {\AA}$).
Although some of the pressure sensitivities seem to be
overestimated, the general behavior is very well predicted as a
direct consequence of the validity of Eq.~\ref{eq:Tcvsra} and the
linear dependence of $\langle r_A \rangle$ to describe the general
behavior of intermediate to large bandwidth manganites in the low
pressure range considered here ($P <$ 1 GPa). For higher pressures,
the linear dependence of $\langle r_A \rangle$ with pressure is no
longer valid, considering the asymptotic behavior of the structural
parameters~\cite{Manolo01}. This fact also explains why our model
predicts a parabolic evolution of $T_c$ with pressure, while the
experimental result in the high pressure range ($P >$ 6 GPa) reveals
an asymptotic behavior~\cite{Postorino03}.

Additionally, in this pressure range, pressure induces distortions
in the MnO$_6$ octahedra~\cite{Manolo01} that can produce a
departure from the expected ideal behavior described by our
empirical model. Finally, we would like to gain insight on the
physical origin of each term in Eq.~\ref{eq:Tcvsra}. As $T_c \sim
W~n$, where $W$ is the bandwidth and $n$ is related to the
electronic density\cite{Millis95}, we may associate $T_c^M$ with
$W$, and the expression between brackets with $n$, that, more
precisely, may represent the relative variation with $\langle r_A
\rangle$ of an effective density of carriers. One possible
association of this $n$ can be established with the $n$ at $T \sim
T_c$ estimated from thermal expansion experiments by Rivadulla et
al.~\cite{Rivadulla06} ($n=1-n_{JT}$, where $n_{JT}$ is the volume
fraction of electrons in the polaronic phase) used to determine the
$T_c[\langle r_A  \rangle (x)$] dependence at constant doping for
the system La$_{2/3}$(Ca$_{1-x}$Sr$_x$)$_{1/3}$MnO$_3$. This implies
that, not only the steric factors that govern the hopping energy and
the polaronic coupling constant that modify the bandwidth will
affect the $T_c$ of the manganite by changing $T_c^M$, but also, and
quantitatively more important, variations on $\langle r_A \rangle$
will determine its value by modifying the effective value of $n$. In
this way, we may associate $T_c^M$ with structural parameters and
with the polaronic narrowing of the bandwidth as~\cite{Medarde95}

\begin{equation}
\label{eq:Tcm} T_c^M \sim W_0~F(E_b) = \frac{cos(w)}{d^{3.5}}~
F(E_b),
\end{equation}

\noindent where $W_0$ is the bare bandwidth, $w$ is the tilt angle
in the plane of the Mn-O bond, $d$ the Mn-O bond length, $E_b$ the
binding energy of polarons and $F$ the appropriate function that
accounts for the polaronic bandwidth reduction.

\begin{figure} [t]
\includegraphics[width=3.3in]{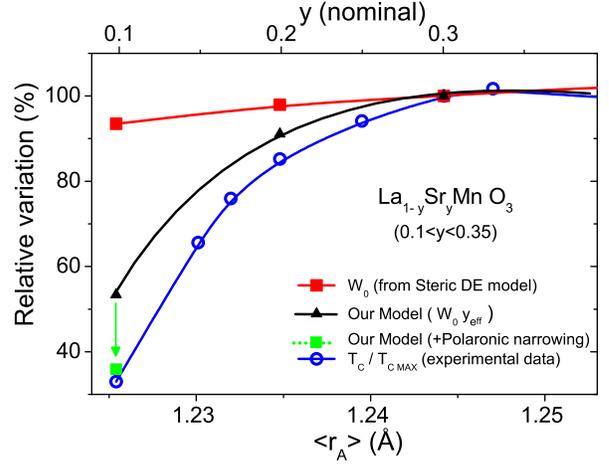}
\vspace{-0mm} \caption{(Color online) The relative variation of the
bare bandwidth $W_0$, of $W_0$ times the term in brackets of
Eq.~\ref{eq:Tcvsra}, of the polaronic correction in the intermediate
to strong coupling regime and of the experimental $T_c$ (from
reference~\cite{Moritomo95}), as a function of the nominal doping
$y$ or of $\langle r_A \rangle$. The arrow shows the correction
added by polaronic effects (only valid for low doping levels).}
\vspace{5mm} \label{fig:relvatc}
\end{figure}

In Fig.~\ref{fig:relvatc} we have plotted the relative variation of
the experimental $T_c$ with $\langle r_A \rangle$ for samples of the
LSMO family. In the same figure we considered the relative variation
of W$_0$, based on the modification of the steric factors. Again, it
is clear that the DE model is far to explain the experimental
variation of $T_c$. If we include the correction of the effective
doping proposed in our phenomenological model a much better
agreement is obtained ($W_0~y_{eff}$). Finally, if we use the
expression $F(E_b)=\exp(\frac{\gamma E_b}{\hbar\omega})$, only valid
for low doping levels as, in this range, we are near the frontier
from strong to intermediate electron-phonon
coupling~\cite{Alex94,Garba04}, we can additionally estimate the
polaronic bandwidth reduction of W$_0$ by using the appropriate
constants~\cite{Deteresa98}. The excellent agreement with the
measured data obtained (marked with an arrow in
Fig.~\ref{fig:relvatc}) indicates that the association of $T_c^M$
with $W$ is a reliable assumption.

\section{CONCLUSIONS}

An empirical law that determines the $T_c$ of T$_{1-y}$D$_y$MnO$_3$
compounds as a function of $\langle r_A \rangle$ and the doping
level $y$ was experimentally extended, using our data and data
already published in the literature for dopings in the 0.15 $\leq y
\leq$ 0.33 range. For these compounds and in the pressure range
where a linear dependence of $\langle r_A \rangle$ with pressure is
still valid, the influence of cationic disorder and the pressure
sensitivity of $T_c$ was quantitatively described by an empirical
relation that associates $T_c$ with the polaronic modified bandwidth
and with an effective doping level, controlled by $\langle r_A
\rangle$.

\section{ACKNOWLEDGEMENTS}

We would like to acknowledge financial support by CONICET Grant PIP
5609 and UBACyT X198. We acknowledge technical assistance from C.
Chiliotte, D. Gim\'enez, E. P\'erez Wodtke and D. Rodr\'{\i}guez
Melgarejo. We are also indebted with A. G. Leyva for providing the
samples and with V. Bekeris for a critical reading of the
manuscript.


\end{document}